**Multi-Block Grid Deformation Method in 3D**

Jie Liu (liu@dixie.edu)
Department of Mathematics
Dixie State College
St George, UT 84770 USA

*Guojun Liao (Corresponding Author: liao@uta.edu)
Department of Mathematics
University of Texas at Arlington
Arlington, TX 76019-0408 USA

**Abstract**: Multi-block grids provide the computational efficiency of structured grids and the flexibility for complex geometry. Thus, Multi-block structured grids are widely used for field simulation on complex domains. In this paper we propose a method which adapts multi-block grids according to a monitor function, which specifies cell volume distribution. The method is an extension of the deformation method on a single block to multi-blocks.

# 1. The Deformation Method

The deformation method for adaptive grids is based on the idea of deforming volume elements of a compact Riemannian manifold [1] [2] by smooth, invertible transformations. In this approach an adaptive grid can be constructed by moving the existing grid nodes with a suitable velocity so that specified cell volume distribution is achieved.

A positive monitor function is defined, which specifies the desired cell volume. It is used to determine a vector field by solving a scalar Poisson equation (if the domain is fixed) or a system consisting of a divergence equation and a curl equation (if the domain is moving). The Appendix contains a detailed description of the method for the fixed domain case.

A main feature of the method is that the moving grid is generated by a transformation whose Jacobian determinant is equal to the given monitor function. As a consequence, grid lines of the same grid family will not cross each other. We now outline the method which solves the following problem:

For a given monitor function $f(x,t) > 0$, find a smooth and convertible transformation $\phi : \Omega_1 \to \Omega_2$ such that $J(\phi(x,t)) = f(\phi(x,t),t)$, where $J(\phi(x,t)) = \det \nabla \phi(x,t)$ is the Jacobian determinant of the transformation.

Solution:

Step 1: Let V = grad (w) where w satisfies the Poisson equation $\Delta \omega = -\dfrac{\partial}{\partial t}\left(\dfrac{1}{f}\right)$ in $\Omega$.

Step 2: Solve the equation below for the transformation $\phi(x,t)$

$$\frac{\partial \phi(x,t)}{\partial t} = f(\phi,t)V(\phi,t) = \eta(\phi,t),$$

with the identity map as the initial condition.

Theorem: $\phi(x,t)$ satisfies, for any t > 0, that $\det \nabla \phi(x,t) = f(\phi,t)$, provided that the equality holds true at t = 0.

This is proved (see [Liu 2006]) by showing that $H = \dfrac{J}{f} = \text{const}$.

Moving grids by the deformation method were generated by the finite difference, the least squares finite element method, and by the streamline stabilization Petrov-Galerkin method, in [3][4][5][7], respectively. In [10], this method is extended to a real time moving grid method for solving time dependent partial differential equations.
In [6] [8], the adaptive deformation method is applied to the Euler equation for compressible fluid flows. A moving grid geometric deformable model using deformation method is developed in [9] for segmentation of medical images. Recently, this method is used to generate adaptive moving grid for particulate flow simulations [15] [16].

The grid deformation method is also applied to the challenging problem of non-rigid image registration. The main idea is to reformulate the variational problem of maximizing a similarity integral as an optimal control problem constrained by the divergence-curl system [17] [18] [19].

## 2. The Multi-Block Method

In [11] [12], a version of the deformation method for multi-block domains in 2D was proposed. In this paper, we extend the method to 3D. The main technical issue is to realize a smooth adaptation across interface of two adjacent blocks. This is accomplished by an iterative procedure designed to match the solution of Poisson equation and its derivative across the interface.

In this section, we describe the method on a two-block domain. Let us consider a 3D back-step and decompose it into two blocks (See Figure 1). The first block (block 1) is the column $[0,1] \times [0,2] \times [0,1]$. A 20×40×20 initial Cartesian uniform grid is generated on it. The second block (block 2) is a cube $[1, 2] \times [0, 1] \times [0, 1]$. The initial Cartesian uniform grid on block 2 is 20×20×20. These two blocks have a common boundary at $x = 1$. Together, these two Cartesian grids form an initial grid on the back-step, which is to be deformed into a moving grid concentrated around a moving sphere of radius $r = 0.2$. The sphere moves from the first block to the second block. Let $l$ denote an artificial time parameter. The coordinate values of the center $(a, b, c)$ of the moving sphere are defined as:

$$\begin{cases} a = 0.5 \\ b = 1.5 - 0.05l \\ c = 0.5 \end{cases} \quad \text{when} \quad 0 \leq l \leq 20$$

$$\begin{cases} a = 0.5 + 0.05(l - 20) \\ b = 0.5 \\ c = 0.5 \end{cases} \quad \text{when} \quad 20 < l \leq 40$$

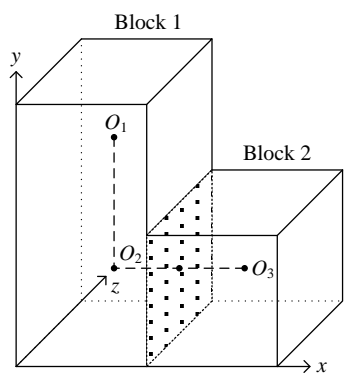

Figure 1 Two blocks of a 3D back-step

At $l = 0$, the center of the sphere is at $O_1(0.5, 1.5, 0.5)$ and then moves to $O_2(0.5, 0.5, 0.5)$ in $0 \le l \le 20$. For $20 < l \le 40$, the center moves into block 2 and reaches $O_3(1.5, 0.5, 0.5)$.

Computation of the moving grid consists of two steps:

Step 1: Let $t$ be another (time) parameter. We deform the initial uniform grid on to a grid adapted to the sphere at $O_1(0.5, 1.5, 0.5)$ in $0 \le t \le 0.5$. $t = 0.5$ corresponds to $l = 0$.

Step 2: For $0 \le l \le 20$, the center of the sphere goes to $O_2(0.5, 0.5, 0.5)$ and for $20 < l \le 40$ the center moves to $O_3(1.5, 0.5, 0.5)$ (See Figure 3.1).

In each step, we define a monitor function on the entire domain through a level set function $d$, which vanishes on the (moving) sphere: $d = (x-a)^2 + (y-b)^2 + (z-c)^2 - r^2$ $(r = 0.2)$.

In step 1, fixing $l = 0$ in $a$, $b$, $c$, for $0 \le t \le 0.5$ we define:

$$\tilde{f} = \begin{cases} 1 & d < -0.05 \\ 1 - 2t + 2t(0.2 - 8d) & -0.05 \le d < 0 \\ 1 - 2t + 2t(0.2 + 8d) & 0 \le d < 0.05 \\ 1 & d \ge 0.05 \end{cases}$$

This is the pre-monitor function before normalization. It is piecewise linear, is small near $d = 0$, and is 1 away from $d = 0$. We then define the normalized monitor function $f$ by

$$f = \frac{\tilde{f} \int_\Omega \frac{1}{\tilde{f}} dA}{|\Omega|}.$$

Note: In step 2, the pre-monitor function is again piecewise linear, is small near $d = 0$, and is 1 away from $d = 0$. Now $d$ depends on the time step $l$. Since the only difference for the two steps is the use of different monitor functions, we will only give details for step 1.

Next, we calculate the right hand side (denoted by RHS) of the Poisson equation

$$\Delta \omega = -\frac{\partial}{\partial t}\left(\frac{1}{f}\right) \text{ in } \Omega$$

and get

$$\text{RHS} = -\frac{\partial}{\partial t}\left(\frac{1}{f}\right) = \frac{\frac{1}{f(x, t+dt)} - \frac{1}{f(x, t)}}{dt}.$$

Now we discretize the Poisson equation by (See Figure 2)

$$\frac{\omega_{i-1,j,k} - 2\omega_{i,j,k} + \omega_{i+1,j,k}}{\Delta x^2} + \frac{\omega_{i,j-1,k} - 2\omega_{i,j,k} + \omega_{i,j+1,k}}{\Delta y^2} + \frac{\omega_{i,j,k-1} - 2\omega_{i,j,k} + \omega_{i,j,k+1}}{\Delta z^2} = rhs_{i,j,k}$$

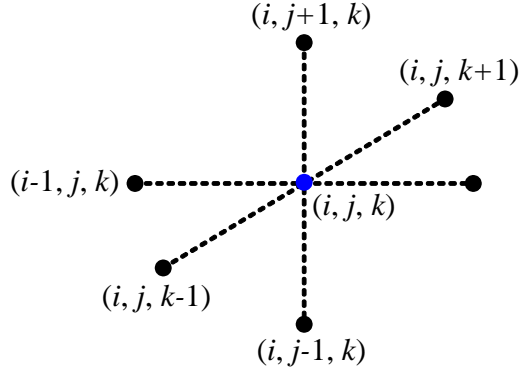

gure 2 Finite difference using adjacent six points

Let $\Delta x = \Delta y = \Delta z = h$. Then for all the interior points we have

$$\omega_{i,j,k} = \frac{1}{6}\left(\omega_{i-1,j,k} + \omega_{i+1,j,k} + \omega_{i,j-1,k} + \omega_{i,j+1,k} + \omega_{i,j,k-1} + \omega_{i,j,k+1} - h^2 \times rhs_{i,j,k}\right).$$

The resulting system of linear algebraic equations is solved by using successive over-relaxation (SOR) method in the following two stages.

$$\tilde{\omega}_{i,j,k} = \frac{1}{6}\left(\omega_{i-1,j,k}^{new} + \omega_{i+1,j,k}^{old} + \omega_{i,j-1,k}^{new} + \omega_{i,j+1,k}^{old} + \omega_{i,j,k-1}^{new} + \omega_{i,j,k+1}^{new} - h^2 \times rhs_{i,j,k}\right)$$

$$\omega_{i,j,k}^{new} = (1-\lambda)\omega_{i,j,k}^{old} + \lambda\tilde{\omega}_{i,j,k}$$

On the boundary, the Neumann boundary condition is implemented. The boundary includes 8 corners, 12 edges and 6 faces for each block. For corner 1 (See Figure 3):

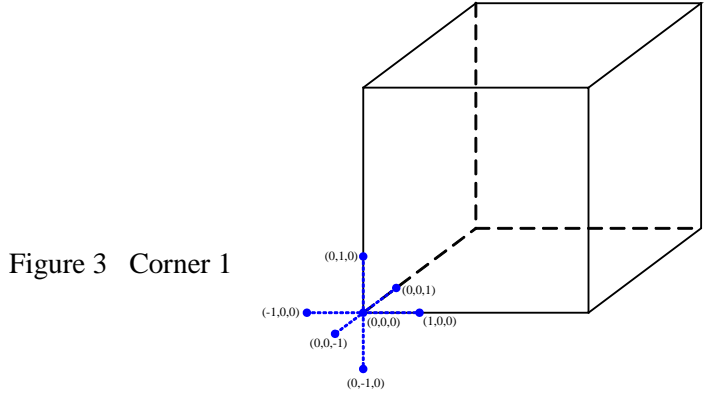

Figure 3  Corner 1

$\dfrac{\partial \omega}{\partial n} = 0$ on $\Gamma$ implies that $\dfrac{\omega_{1,0,0} - \omega_{-1,0,0}}{2\Delta x} = 0 \Rightarrow \omega_{1,0,0} = \omega_{-1,0,0}$

$\dfrac{\omega_{0,1,0} - \omega_{0,-1,0}}{2\Delta y} = 0 \Rightarrow \omega_{0,1,0} = \omega_{0,-1,0}$

$\dfrac{\omega_{0,0,1} - \omega_{0,0,-1}}{2\Delta z} = 0 \Rightarrow \omega_{0,0,1} = \omega_{0,0,-1}$.

Thus, we have

$$\omega_{0,0,0} = \frac{1}{6}\left(\omega_{-1,0,0} + \omega_{1,0,0} + \omega_{0,-1,0} + \omega_{0,1,0} + \omega_{0,0,-1} + \omega_{0,0,1} - h^2 \times rhs_{0,0,0}\right)$$

$$= \frac{1}{6}\left(2\omega_{1,0,0} + 2\omega_{0,1,0} + 2\omega_{0,0,1} - h^2 \times rhs_{0,0,0}\right)$$

Other corner points are treated in a similar way.
For edge 1 (See Figure 4):

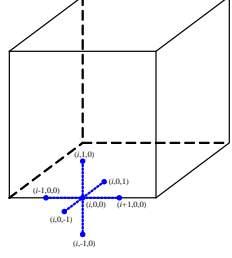

Figure 4 Edge 1

$\dfrac{\partial \omega}{\partial n} = 0$ on $\Gamma$ again implies that

$$\dfrac{\omega_{i,1,0} - \omega_{i,-1,0}}{2\Delta y} = 0 \Rightarrow \omega_{i,1,0} = \omega_{i,-1,0}$$

$$\dfrac{\omega_{i,0,1} - \omega_{i,0,-1}}{2\Delta z} = 0 \Rightarrow \omega_{i,0,1} = \omega_{i,0,-1}$$

$$\omega_{i,0,0} = \dfrac{1}{6}\left(\omega_{i-1,0,0} + \omega_{i+1,0,0} + \omega_{0,-1,0} + \omega_{0,1,0} + \omega_{0,0,-1} + \omega_{0,0,1} - h^2 \times rhs_{0,0,0}\right)$$

$$= \dfrac{1}{6}\left(\omega_{i-1,0,0} + \omega_{i+1,0,0} + \omega_{0,1,0} + \omega_{0,1,0} + \omega_{0,0,1} + \omega_{0,0,1} - h^2 \times rhs_{0,0,0}\right)$$

$$= \dfrac{1}{6}\left(\omega_{i-1,0,0} + \omega_{i+1,0,0} + 2\omega_{0,1,0} + 2\omega_{0,0,1} - h^2 \times rhs_{i,0,0}\right)$$

Other boundary edges are treated in the similar way.
Similarly, for face 1 (See Figure 5), we have:

$$\dfrac{\omega_{i,1,j} - \omega_{i,-1,j}}{2\Delta y} = 0 \Rightarrow \omega_{i,1,j} = \omega_{i,-1,j}$$

$$\omega_{i,0,j} = \dfrac{1}{6}\left(\omega_{i-1,0,j} + \omega_{i+1,0,j} + \omega_{i,-1,j} + \omega_{i,1,j} + \omega_{i,0,j-1} + \omega_{i,0,j+1} - h^2 \times rhs_{i,0,j}\right)$$

$$= \dfrac{1}{6}\left(\omega_{i-1,0,j} + \omega_{i+1,0,j} + \omega_{i,1,j} + \omega_{i,1,j} + \omega_{i,0,j-1} + \omega_{i,0,j+1} - h^2 \times rhs_{i,0,j}\right)$$

$$= \dfrac{1}{6}\left(\omega_{i-1,0,j} + \omega_{i+1,0,j} + 2\omega_{i,1,j} + \omega_{i,0,j-1} + \omega_{i,0,j+1} - h^2 \times rhs_{i,0,j}\right)$$

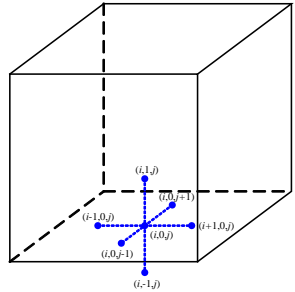

Figure 5 Face 1

Other boundary faces are treated in the similar way.

Special care should be taken for the interior points of the interface between the blocks. Neumann boundary condition does not apply on these points. We denote the points in block 1 by $A_{i,j,k}$, points of block 2 by $B_{i,j,k}$, respectively. For the points on the interface, we use points from both blocks. For example, at x = 1, i = mx = 20, $1 < j < my - 1, 1 < k < mz - 1$, where mx = my = mz = 20, we have, for block 1:

$$\omega A_{mx,j,k} = \frac{1}{6}\left(\omega A_{mx-1,j,k} + \omega B_{1,j,k} + \omega A_{i,j-1,k} + \omega A_{i,j+1,k} + \omega A_{i,j,k-1} + \omega A_{i,j,k+1} - h^2 \times rhs_{i,j,k}\right).$$

For block 2:

$$\omega B_{0,j,k} = \frac{1}{6}\left(\omega A_{mx-1,j,k} + \omega B_{0,j,k} + \omega B_{i,j-1,k} + \omega B_{i,j+1,k} + \omega B_{i,j,k-1} + \omega B_{i,j,k+1} - h^2 \times rhs_{i,j,k}\right).$$

The 4 corners and 4 edges on the interface are implemented in the similar way:

$$\omega A_{mx,0,0} = \frac{1}{6}\left(2\omega A_{mx,0,0} + \omega B_{1,0,0} + \omega A_{mx-1,0,0} + \omega A_{mx,1,0} + \omega A_{mx,0,1} - h^2 \times rhs_{mx,0,0}\right)$$

$$\omega B_{0,0,0} = \frac{1}{6}\left(2\omega B_{0,0,0} + \omega A_{mx-1,0,0} + \omega B_{1,0,0} + \omega B_{0,1,0} + \omega B_{0,0,1} - h^2 \times rhs_{0,0,0}\right)$$

$$\omega A_{mx,0,k} = \frac{1}{6}\left(\omega A_{mx-1,0,k} + \omega B_{1,0,k} + \omega A_{mx,0,k+1} + \omega A_{mx,0,k-1} + \omega A_{mx,1,k} + \omega A_{mx,0,k} - h^2 \times rhs_{mx,0,0}\right)$$

$$\omega B_{0,0,k} = \frac{1}{6}\left(\omega B_{1,0,k} + \omega A_{mx,0,k} + \omega B_{0,0,k+1} + \omega B_{0,0,k-1} + \omega B_{0,1,k} + \omega B_{0,0,k} - h^2 \times rhs_{mx,0,0}\right)$$

Here, $\omega A$ denotes values of $\omega$ in block 1, $\omega B$ denotes values in block 2. Other points are implemented similarly.

After solving for $\omega$ from the Poisson equation, we then compute its gradient field $V = \nabla \omega$. The nodal velocity field is then defined by $\eta = f V$.

At last, the new position of any grid point is computed by the following ordinary differential equation:

$$\frac{\partial \phi}{\partial t} = f V, \ t>0, \text{ with } \phi(0) = \text{the identity map}.$$

Next, we present a numerical example to demonstrate the method.

## 3. A Numerical Example

The computed results are shown in Figure 6a to Figure 14b. A "deformed sphere" is formed and moves from the (red) block with $0 \le x \le 1$ to the (green) block with $1 \le x \le 2$.

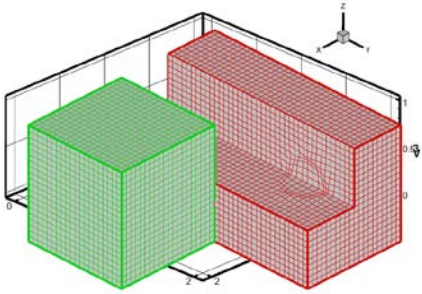
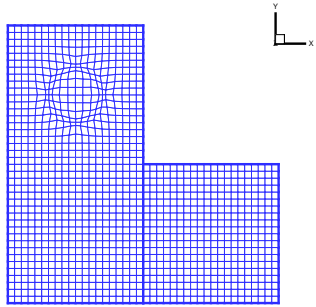

Figure 6a Cutaway plot at l = 0          Figure 6b Slice at l = 0

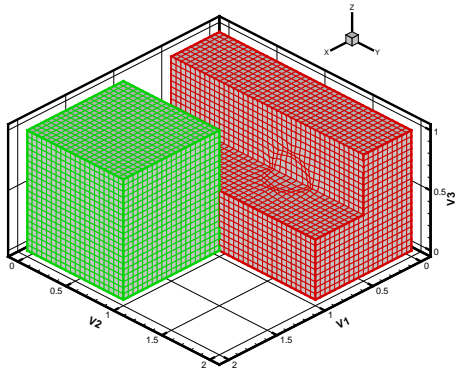 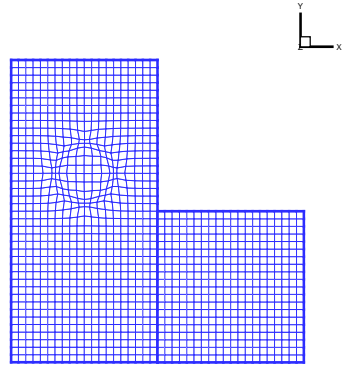

Figure 7a Cutaway plot at l = 5    Figure 7b Slice at $z = 0.5$ at l = 5

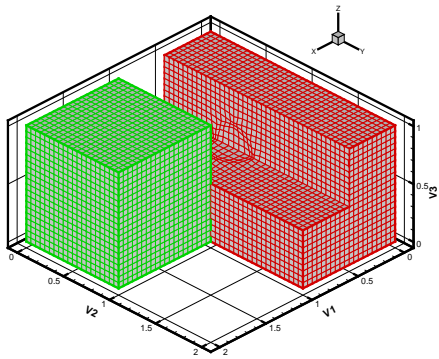 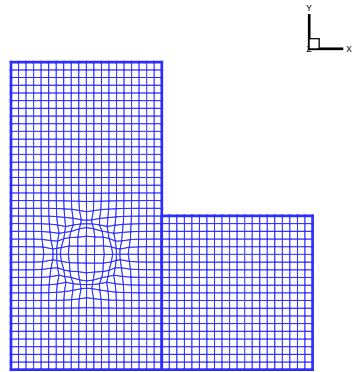

Figure 8a Cutaway plot at l = 15   Figure 8b Slice at $z = 0.5$ at l = 15

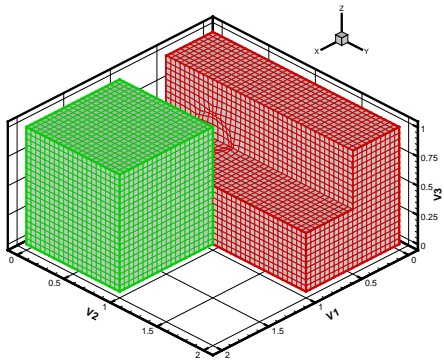 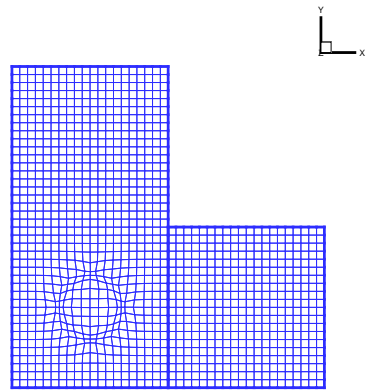

Figure 9a Cutaway plot at l = 20   Figure 9b Slice at $z = 0.5$ at l = 20

Remark 1: So far, the "deformed sphere" has moved inside the block 1, maintaining the desired cell size distribution. In next several steps, the "deformed sphere" moves towards the interface of the two blocks.

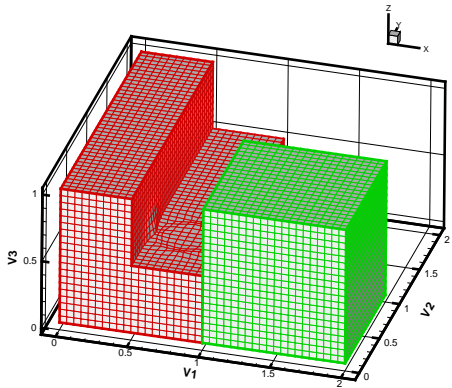 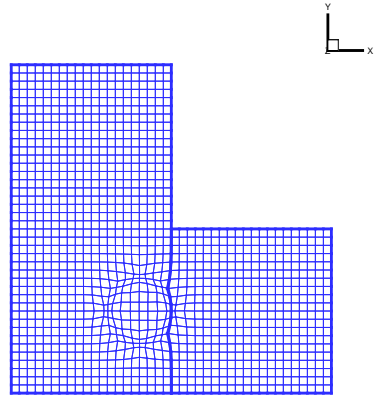

Figure 10a Cutaway plot at l = 25    Figure 10b Slice at $z = 0.5$ at l = 25

Remark 2: At this time step l = 25, the "deformed sphere" reaches the block interface with x = 1. Our method assures smooth crossing through the interface, maintaining the desired cell size distribution.

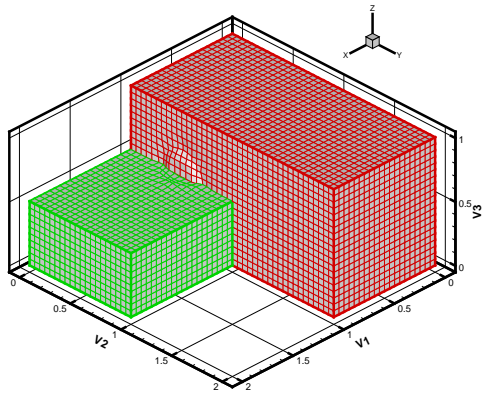 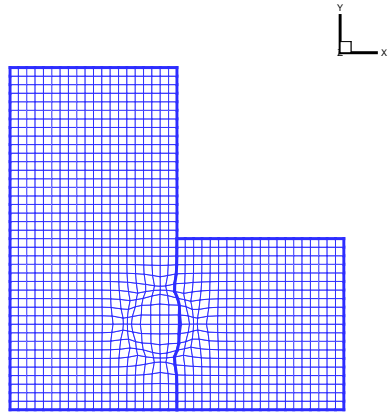

Figure 11a Cutaway plot for time step l = 28    Figure 11b Slice at $z = 0.5$ at l = 28

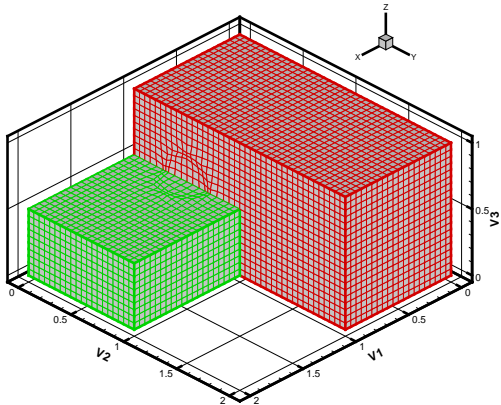 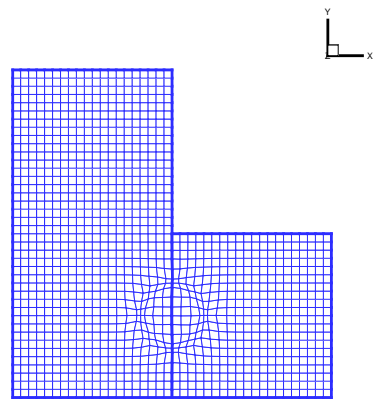

Figure 12a Cutaway plot at l = 30    Figure 12 b Slice at $z = 0.5$ at l = 30

Remark 3: At this time step, the "deformed sphere" is located between the two blocks. Our method assures smooth crossing through the interface, maintaining the desired cell size distribution.

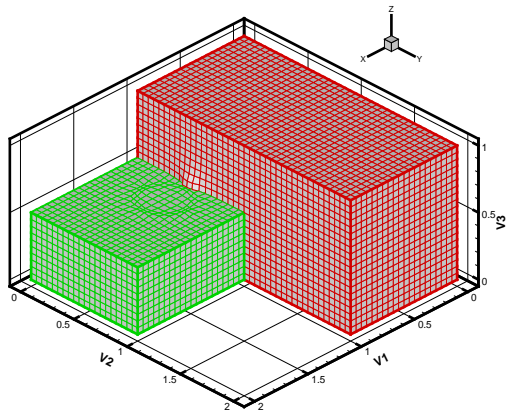
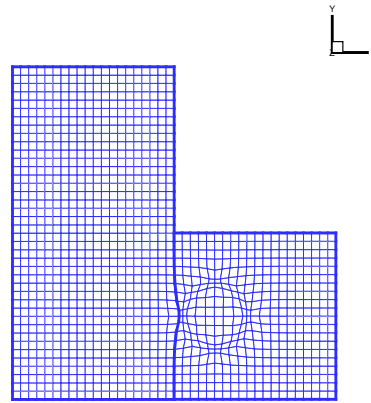

Figure 13a Cutaway plot for time step l = 35    Figure 13b Slice at $z = 0.5$ at l = 35

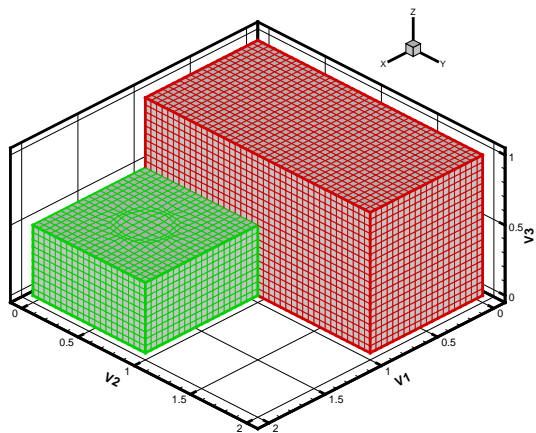
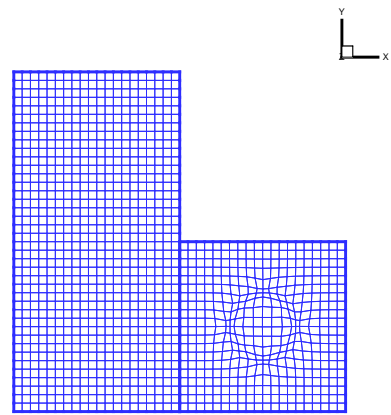

Figure 14a Cutaway plot at l = 40    Figure 14b Slice at $z = 0.5$ at l = 40

## 4. Conclusions

A three dimensional multi-block deformation method is proposed and implemented in a two-block domain. Our 3D numerical example shows that the "deformed sphere" passes through the interface between the two blocks smoothly.
The implementation on domains with more blocks can be done in a similar way as in this two-block example. Parallel computing can be used to concurrently calculate the grids on different blocks.
The fist author is grateful to The University of Texas at Arlington for financial support in her Ph D study. The second author is grateful for support of the National Science Foundation of United States through its Computational Mathematics Program in the Division of Mathematical Sciences. The opinions expressed in this paper are those of the Authors.